\documentclass[fleqn,twoside]{article}
\usepackage[headings]{espcrc2}

\readRCS
$Id: espcrc2.tex,v 1.2 2004/02/24 11:22:11 spepping Exp $
\usepackage{epsfig}
\usepackage[figuresright]{rotating}

\def\d{{\rm d}}

\def\beq{\begin{equation}}
\def\eeq{\end{equation}}
\def\bea{\begin{eqnarray}}
\def\eea{\end{eqnarray}}

\title{Polarization and Resummation in Slepton Production at Hadron Colliders}

\author{M. Klasen\address{Laboratoire de Physique Subatomique et de
 Cosmologie, Universit\'e Joseph Fourier/CNRS-IN2P3, \\
 53 Avenue des Martyrs, F-38026 Grenoble, France}}

\runtitle{Polarization and Resummation in Slepton Production at Hadron Colliders}
\runauthor{M. Klasen}

\begin{document}

\begin{abstract}
In $R$-parity conserving supersymmetric (SUSY) models, sleptons are produced
in pairs at hadron colliders through neutral and charged electroweak
currents. We demonstrate that the polarization of the initial hadron beams
allows for a direct extraction of the slepton mixing angle and thus for a
determination of the underlying SUSY-breaking mechanism. We also perform a
first precision calculation of the transverse-momentum ($q_T$) spectrum of
the slepton pairs by resumming soft multiple-gluon emission at
next-to-leading logarithmic order. The results show a relevant contribution
of resummation both in the small and intermediate $q_T$-regions, which
strongly influences the extraction of the missing transverse-momentum
signal and the subsequent slepton mass-determination, and little dependence
on unphysical scales and non-perturbative contributions.
\vspace{1pc}
\end{abstract}

\maketitle

\vspace*{-9cm}\noindent LPSC 06-33
\vspace*{ 8cm}

\section{Introduction}

The Minimal Supersymmetric Standard Model (MSSM) \cite{Nilles:1983ge,%
Haber:1984rc} is one of the most promising extensions of the Standard Model
(SM) of particle physics. It postulates a symmetry between fermionic and
bosonic degrees of freedom in nature and predicts the existence of a
fermionic (bosonic) supersymmetric (SUSY) partner for each bosonic
(fermionic) SM particle. It provides a qualitative understanding of various
phenomena in particle physics, as it stabilizes the gap between the Planck
and the electroweak scale, leads to gauge coupling
unification in a straightforward way, and includes
the lightest supersymmetric particle as a dark matter candidate.
Therefore the search for supersymmetric particles is
one of the main topics in the experimental program of present (Fermilab
Tevatron) and future (CERN LHC) hadron colliders.

SUSY must be broken at low energy, since spin partners of the SM particles
have not yet been observed. As a consequence, the squarks, sleptons,
charginos, neutralinos and gluino of the MSSM must be massive in comparison
to their SM counterparts. The LHC will perform a conclusive search covering
a wide range of masses up to the TeV scale. Total production cross sections
for SUSY particles at hadron colliders have been extensively studied in the
past at leading order (LO) \cite{Dawson:1983fw,delAguila:1990yw,Baer:1993ew}
and also at next-to-leading order (NLO) of perturbative QCD
\cite{Beenakker:1996ch,Beenakker:1997ut,Berger:1998kh,Berger:1999mc,%
Berger:2000iu,Baer:1997nh,Beenakker:1999xh}.

\section{LO cross section with polarization}

Despite the first successful runs of the RHIC collider in the polarized $pp$
mode (and due to its limited energy range of $\sqrt{S}\leq 500$ GeV),
polarized SUSY production cross sections have received much less attention.
Only the pioneering LO calculations for massless squark and gluino
production \cite{Craigie:1983as,Craigie:1984tk} have recently been
confirmed, extended to the massive case, and applied to current hadron
colliders \cite{Gehrmann:2004xu}.

Due to their purely electroweak couplings, sleptons are among the lightest
SUSY particles in many SUSY-breaking scenarios \cite{Allanach:2002nj}.
Sleptons and sneutrinos often decay directly into the stable lightest SUSY
particle (lightest neutralino in mSUGRA models or gravitino in GMSB models)
plus the corresponding SM partner (lepton or neutrino). As a result, the
slepton signal at hadron colliders will consist in a highly energetic lepton
pair, which will be easily detectable, and associated missing energy.

The neutral current cross section for the production of non-mixing slepton
pairs in collisions of quarks with definite helicities has been calculated
in \cite{Chiappetta:1985ku}. In general SUSY-breaking models, where the
sfermion interaction eigenstates are not identical to the respective mass
eigenstates, the left- and right-handed coupling strengths must be
multiplied by $S_{j1}S_{i1}^\ast$ and $S_{j2} S_{i2}^\ast$ ($i,j=L,R$),
respectively, where the unitary matrix $S$ diagonalizes the sfermion mass
matrix, $S {\mathcal M}^2 S^\dagger={\rm diag}\, (m_1^2,m_2^2)$. These
slepton mixing effects have recently been included in the polarized
hadroproduction cross section \cite{Bozzi:2004qq}. One can then calculate
the longitudinal double-spin asymmetry, however with the result that
\bea
 A_{LL}\!\!\! &=& \!\!\!{\d\Delta\hat{\sigma}_{LL}\over\d\hat{\sigma}} = -1
\eea
is totally independent of all SUSY-breaking parameters. It is thus far more
interesting to calculate the single-spin asymmetry $A_L=\d\Delta
\hat{\sigma}_L/\d\hat{\sigma}$ from the polarized differential cross section
\bea
 \d\Delta\hat{\sigma}_L\!\!\!&=&\!\!\!
 {\d\hat{\sigma}_{ 1, 1}
 +\d\hat{\sigma}_{ 1,-1}
 -\d\hat{\sigma}_{-1, 1}
 -\d\hat{\sigma}_{-1,-1}
 \over 4},
\eea
{\it i.e.} for the case of only one polarized hadron beam. Not only does the
neutral current cross section
remain sensitive to the SUSY-breaking parameters, but even more the
squared photon contribution, which is insensitive to these parameters, is
eliminated. Finally, this scenario may also be easier to implement
experimentally, {\it e.g.} at the Tevatron, since protons are much more
easily polarized than antiprotons \cite{Baiod:1995eu}.

For the only existing polarized hadron collider RHIC, which will be
operating at a center-of-mass energy of $\sqrt{S}=500$ GeV in the near
future, and in the GMSB model with a light tau slepton, we show the
single-spin asymmetry in Fig.\ \ref{fig:2} as a function of the cosine of
%
\begin{figure}
 \centering
 \epsfig{file=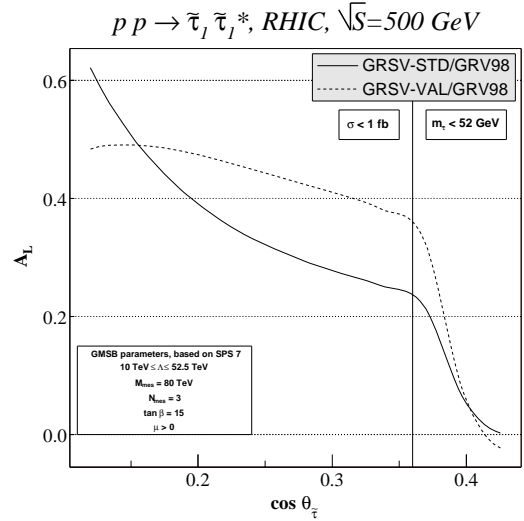,width=\columnwidth}
 \caption{\label{fig:1}Dependence of the longitudinal single-spin asymmetry
 $A_L$ on the cosine of the stau mixing angle for $\tilde{\tau}_1$ pair
 production in a GMSB model at RHIC.}
\end{figure}
%
the stau mixing angle. The
asymmetry is quite large and depends strongly on the stau mixing angle.
However, very large values of $\cos\theta_{\tilde{\tau}}$ and stau masses
below 52 GeV may already be excluded by LEP \cite{Barate:1998zp}, while
small values of $\cos\theta_{\tilde{\tau}}$ may be unaccessible at RHIC due
to its limited luminosity, which is not expected to exceed 1 fb$^{-1}$.
Polarization of the proton beam will also not be perfect, and the calculated
asymmetries should be multiplied by the degree of beam polarization $P_L
\simeq 0.7$. The uncertainty introduced by the polarized parton densities
increases considerably to the left of the plot, where the stau mass 41 GeV
$\leq m_{\tilde{\tau}}\leq$ 156 GeV and the associated values of the parton
momentum fractions $x_{a,b}\simeq2m_{\tilde{\tau}}/\sqrt{S}$ become large.

The SM background cross section can be reduced by imposing an invariant
mass cut on the observed tau lepton pair, {\it e.g.}
of 2$\cdot$52 GeV. While the cross section of 0.13 pb is then still two
orders of magnitude larger than the SUSY signal cross section of 1 fb, the
SM asymmetry of -0.04 for standard polarized parton densities or -0.10 for
the valence-type polarized parton densities can clearly be distinguished
from the SUSY signal due to its different sign.

\section{Total cross section at NLO}

The QCD corrections to the total slepton-pair production cross section
in quark-antiquark annihilation are completely equivalent to those
encountered in the Drell-Yan process and have been calculated (for the
unpolarized case) in Ref.~\cite{Baer:1997nh}. For a complete analysis that
is consistent at ${\cal O}(\alpha_s)$, also the SUSY-QCD corrections must be
added, which affect the $q\bar{q}V$ vertices \cite{Beenakker:1999xh}. Since
heavy-mass SUSY particles (squarks and gluinos) are involved in the loops,
the genuine SUSY corrections are expected to be considerably smaller than
the standard QCD corrections.

\begin{figure}
 \centering
 \epsfig{file=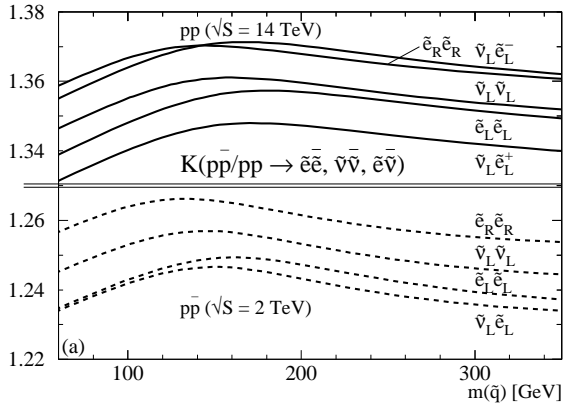,width=\columnwidth}
 \caption{\label{fig:2}Ratios of the total NLO SUSY-QCD cross sections to
 the corresponding LO cross sections for slepton-pair production at the LHC
 (full) and Tevatron (dashed) as a function of the squark mass
 $m_{\tilde{q}}$ for a gluino mass of 200~GeV and slepton masses of
 ${\tilde{e}_R}/{\tilde{e}_L}/{\tilde{\nu}_L} = 120/150/135$ GeV.}
\end{figure}

This is indeed borne out by detailed calculations, an example of which is
presented in Fig.~\ref{fig:2}. The SUSY-QCD $K$-factors differ little from
%
%
the QCD $K$-factors, which are approached in the asymptotic limit of
large $\tilde{q}/\tilde{g}$ masses at the right-hand $y$-axis of the
figure. The QCD corrections to the production of ${\tilde \mu}$ and ${\tilde
\tau}$ pairs follow the same pattern for equivalent invariant masses
of the pairs.

\section{Transverse-momentum distribution at NLL}

Since in hadronic collisions the longitudinal momentum balance is unknown,
a precise knowledge of the transverse-momentum ($q_T$) balance is of vital
importance for the discovery of SUSY particles. In the case of sleptons,
the Cambridge (s)transverse mass $m_{T2}$ proves to be particularly useful
for the reconstruction of their masses \cite{Lester:1999tx} and
determination of their spin \cite{Barr:2005dz}, the two key features that
distinguish them from SM leptons produced mainly in $WW$ or $t\bar{t}$
decays \cite{Lytken:22,Andreev:2004qq}. Furthermore, both detector
kinematic acceptance and efficiency depend, of course, on $q_T$.

When studying the $q_T$-distribution of a slepton pair produced with
invariant mass $M$ in a hadronic collision, it is appropriate to separate
the large-$q_T$ and small-$q_{T}$ regions. In the
large-$q_T$ region ($q_T\geq M$) the use of fixed-order perturbation theory
is fully justified, since the perturbative series is controlled by a small
expansion parameter, $\alpha_s(M^2)$.
The bulk of the events will be produced in the small-$q_{T}$ region, where
the coefficients of the perturbative expansion in $\alpha_s(M^{2})$ are
enhanced by powers of large logarithmic terms, $\ln(M^{2}/q_{T}^{2})$. As
a consequence, results based on fixed-order calculations diverge as $q_T
\to 0$, and the convergence of the perturbative series is spoiled. These
logarithms are due to multiple soft-gluon emission from the initial state
and have to be systematically resummed to all orders in $\alpha_s$ in
order to obtain reliable perturbative predictions. The method to perform
all-order soft-gluon resummation at small $q_{T}$ is well known
\cite{Bozzi:2005wk}. At intermediate $q_T$ the resummed result has to be
consistently matched with fixed-order perturbation theory in order to
obtain predictions with uniform theoretical accuracy over the entire range
of transverse momenta.

We have recently implemented the formalism proposed in \cite{Bozzi:2005wk}
and computed the $q_T$-distribution of a slepton pair produced at the LHC by
combining NLL resummation at small $q_T$ and LO (${\cal O}(\alpha_s)$)
perturbation theory at large $q_T$ \cite{Bozzi:2006fw}. In this computation,
we use the MRST (2004) NLO set of parton distribution functions
\cite{Martin:2004ir} and $\alpha_s$ evaluated at two-loop accuracy. We fix
the resummation scale $Q$ equal to the invariant mass $M$ of the slepton
(slepton-sneutrino) pair and we allow $\mu=\mu_{F}=\mu_{R}$ to vary between
$M/2$ and $2M$ to estimate the perturbative uncertainty. We also integrate
the double-differential cross section d$\sigma/$d$M^2$d$q_T^2$ with respect
to $M^{2}$, taking as lower limit the energy threshold for $\tilde{\tau}_{1}
\tilde{\tau}_{1}^{*} (\tilde{\tau}_1\tilde{\nu}_{\tau})$ production and as
upper limit the hadronic energy ($\sqrt S$=14 TeV at the LHC).

\begin{figure}
 \centering
 \epsfig{file=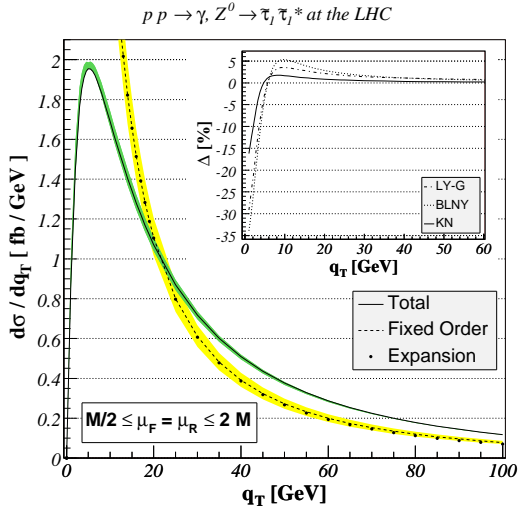,width=\columnwidth}
 \caption{\label{fig:3} Transverse-momentum distribution of
 $\tilde{\tau}_1$-pairs at the LHC at LO (dashed), NLL+LO (full) and after
 asymptotic expansion of the resummation formula (dotted). The shaded bands
 demonstrate the respective scale uncertainties, while the insert shows
 three estimates of non-perturbative contributions.}
\end{figure}

\begin{figure}
 \centering
 \epsfig{file=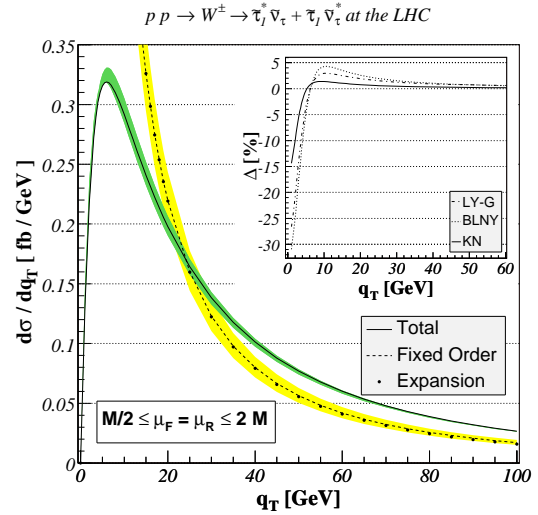,width=\columnwidth}
 \caption{\label{fig:4} Same as Fig.\ \ref{fig:3} for the process $pp\to
 \tilde{\tau}_1\tilde{\nu}_{\tau}^*+\tilde{\tau}_1^*\tilde{\nu}_{\tau}$.}
\end{figure}

In the case of $\tilde{\tau}_1 \tilde{\tau}_1^*$ production (neutral
current process, see Fig.\ \ref{fig:3}), we choose the SPS7
%
%
mSUGRA benchmark point \cite{Allanach:2002nj} which gives, after the
renormalization group (RG) evolution of the SUSY-breaking parameters
performed by the SUSPECT computer program \cite{Djouadi:2002ze}, a light
${\tilde \tau}_{1}$ of mass $m_{{\tilde \tau}_{1}}=114$ GeV.   
In the case of $\tilde{\tau}_1\tilde{\nu}_{\tau}^*+\tilde{\tau}_1^*
\tilde{\nu}_{\tau}$ production (charged current process, see Fig.\
\ref{fig:4}), we use instead the SPS1 mSUGRA benchmark point
%
%
which gives a light ${\tilde \tau}_{1}$ of mass $m_{{\tilde \tau}_{1}}=136$
GeV as well as a light $\tilde{\nu}_{\tau}$ of mass $m_{{\tilde \nu}_{\tau}}
=196$ GeV.   

In both cases we plot the LO result (dashed line), the expansion of the
resummation formula at LO (dotted line), the total NLL+LO matched result
(solid line), the uncertainty band from scale variation, and the
quantity
\begin{equation}
 \Delta = \frac{d\sigma^{\rm (res.+NP)}(\mu=M)-d\sigma^{(\rm res.)}(\mu=M)}
 {d\sigma^{(\rm res.)} (\mu=M)}.
\end{equation}
The parameter $\Delta$ gives thus an estimate of the contributions from
different NP parameterizations (LY-G \cite{Ladinsky:1993zn}, BLNY
\cite{Landry:2002ix}, KN \cite{Konychev:2005iy}) that we included in
the resummed formula.

We can see that the LO result diverges to $+ \infty$, as expected, for both
processes as $q_{T}\to 0$, and the asymptotic expansion of the resummation
formula at LO is in very good agreement with LO both at small and
intermediate values of $q_{T}$. The effect of resummation is clearly visible
at small and intermediate values of $q_T$, the resummation-improved result
being nearly 39\% (36\%) higher at $q_T=50$ GeV than the pure fixed order
result in the neutral (charged) current case.
When integrated over $q_T$, the former leads to a total cross section of
66.8 fb (12.9 fb) in good agreement (within 3.5\%) with the QCD-corrected
total cross section at ${\cal O}(\alpha_s)$ \cite{Beenakker:1999xh}.

The scale dependence is clearly improved in both cases with respect to the
pure fixed-order calculations. In the small and
intermediate $q_{T}$-region (up to 100 GeV) the effect of scale variation is
~10\% for the LO result, while it is always less than 5\% for the NLL+LO
curve.
Finally, non-perturbative contributions are under good control. Their effect
is always less than 5\% for $q_{T} >$ 5 GeV and thus considerably smaller
than resummation effects.

\section{Conclusions}
The recent luminosity upgrade of the Tevatron and the imminent start-up of
the LHC put the discovery of SUSY particles with masses up to the TeV-scale
within reach. Since this discovery depends critically on the
missing transverse-energy signal, a precise knowledge not only of the total
cross section, but also of the $q_T$-spectrum is mandatory. We have
performed a first step in this direction by resumming multiple soft-gluon
emission for slepton-pair production at next-to-leading logarithmic order
and matching it to the fixed-order calculation. We demonstrated that the
total cross section at NLO is reproduced very well. (S)transverse mass
measurements will then lead to precise slepton mass and spin determinations,
while polarization of the initial hadron beams may allow for an extraction
of the slepton mixing angle and the underlying SUSY-breaking parameters.

\end{document}